\documentclass[journal]{IEEEtran}

\usepackage{cite}
\usepackage[pdftex]{graphicx}
\usepackage{amsmath}
\usepackage{url}

\begin{document}

\title{Aqua MODIS Band 24 Crosstalk Striping}

\author{Graziela~R.~Keller,~Zhipeng~Wang,~Aisheng~Wu,~and~Xiaoxiong~Xiong
\thanks{Manuscript received XXXX XX, 2016; revised XXXX XX, 2016.}
\thanks{G. R. Keller, Z. Wang, and A. Wu  are with Science Systems and Applications, Inc., Lanham, MD, 20706 USA (e-mail: graziela.keller@ssaihq.com, zhipeng.wang@ssaihq.com, and aisheng.wu@ssaihq.com)}
\thanks{X. Xiong is with Sciences and Exploration Directorate, NASA Goddard Space  Flight  Center,  Greenbelt,  MD  20771  USA  (e-mail:  xiaoxiong-1@nasa.gov)}}

\markboth{IEEE GEOSCIENCE AND REMOTE SENSING LETTERS,~VOL.~X, No.~X, XXXX~2016}%
{Keller \MakeLowercase{\textit{et al.}}: Aqua MODIS Band 24 Crosstalk Striping}

\maketitle

\begin{abstract}
Aqua MODIS, unlike its predecessor on board the Terra spacecraft, had always been thought to have been spared from significant deleterious impacts of electronic crosstalk on its imagery.
However, recent efforts brought to our attention the presence of striping artifacts in Aqua MODIS images from band 24 (4.47$\mu$m), which upon further inspection proved to have a noticeable impact 
on the quality of the L1B product and to have been 
present since the beginning of the mission, in 2002. Using images of the Moon from scheduled lunar observations, 
we linked the artifacts with electronic crosstalk contamination of the response of detector 1 of band 24 by 
signal sent from the detector 10 of band 26 (1.375$\mu$m), a neighboring band in the same focal plane assembly. In this paper, we report on these findings,
the artifact mitigation strategy adopted by us, and on our success in restoring band 24 detector 1 behavior and image quality.

\end{abstract}


\IEEEpeerreviewmaketitle

\section{Introduction}
\IEEEPARstart{T}{he} Moderate Resolution Imaging Spectroradiometer (MODIS) on board the low orbit platform Aqua was launched on May 4, 2002. It is a cross-track 
scanner that uses a rotating mirror to image the Earth in 16 medium- and long-wave infrared (MWIR, LWIR) thermal emissive bands (TEB) 
spanning a 3.7 -- 14.4 $\mu$m wavelength interval and 20 visible, near-, and short-wave infrared (VIS, NIR, and SWIR) reflective solar bands (RSB) covering the 0.4 -- 2.2 $\mu$m interval. 
Each of the MODIS' bands comprises its own array of 
detectors, each array being placed in one of 4 separate focal plane assemblies (FPA), according to the wavelength regime covered by the bands. 
Within each focal plane, the arrays of detectors corresponding to different bands are placed side-by-side, in parallel rows, in the along-scan (cross-track) direction, sequentially imaging the same Earth target, as the mirror rotates \cite{xiong2009overview}.

\begin{figure}[!h]
\centering
\includegraphics[width=3.5in,height=2in]{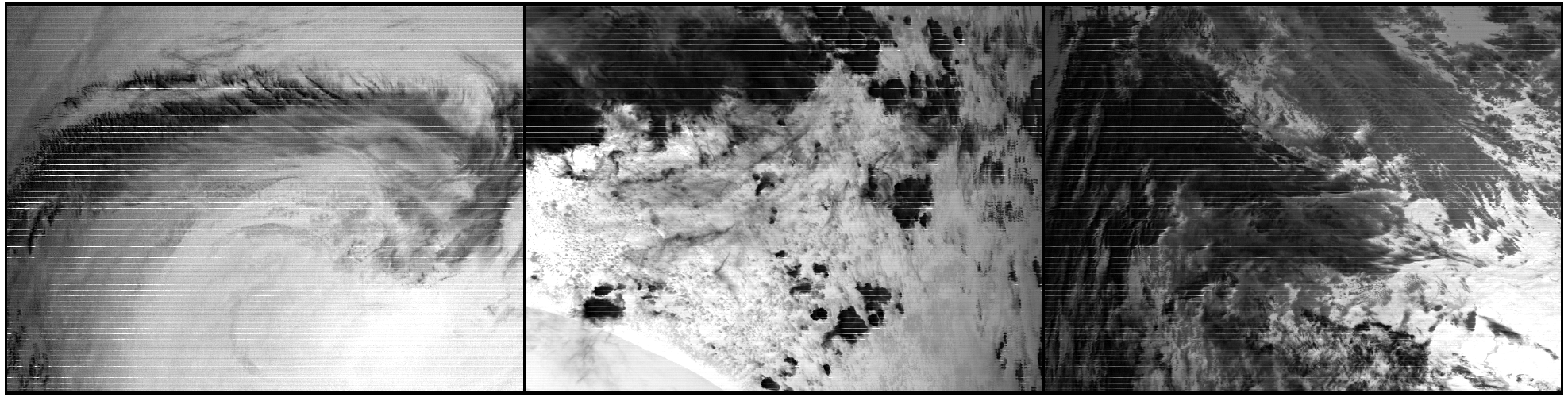}
\caption{Sample L1B images from Aqua MODIS band 24, from 2016 granules, showing striping. Here and throughout the paper, darker regions in the imagery correspond to lower signal. 
The grayscale colormap of each of the three images is scaled independently.}
\label{fig_striping}
\end{figure}

Distinct striping artifacts are evident in Aqua MODIS images from band 24 - a TEB covering the 4.433 -- 4.498 $\mu$m wavelength interval with a 
typical radiance of 0.17 W/m$^{2}$/$\mu$m/sr, corresponding to a typical brightness temperature (T$_{typ}$) of 250 K.
Band 24, designed to measure atmospheric temperature, is mounted on the SWIR/MWIR FPA and its detectors are arranged into a 10$\times$1 array, identified by numbers between 1 and 10, 
in a sequence referred to as the ``product order" \cite{xiong2009aqua}.  
Sample Aqua band 24 L1B images affected by striping are shown in Fig. \ref{fig_striping}.  

All ten Aqua MODIS band 24 detectors are currently flagged as operational by the MODIS Characterization Support Team (MCST), and have historically shown 
stable noise-equivalent temperature differences (NEdT) well below the 0.25 K requirement for this band. 

In this paper, we describe how we identified and characterized Aqua MODIS band 24 striping artifacts, 
how we tracked their source as being crosstalk contamination from a neighboring RSB using scheduled Moon observations, 
and finally, we describe the mitigation strategies we implemented and discuss their impact on the L1B product.       

\section{Methodology}
\subsection{Characterization of band 24 Striping}

\begin{figure}[!t]
\centering
\includegraphics[width=3.5in,height=1.6in]{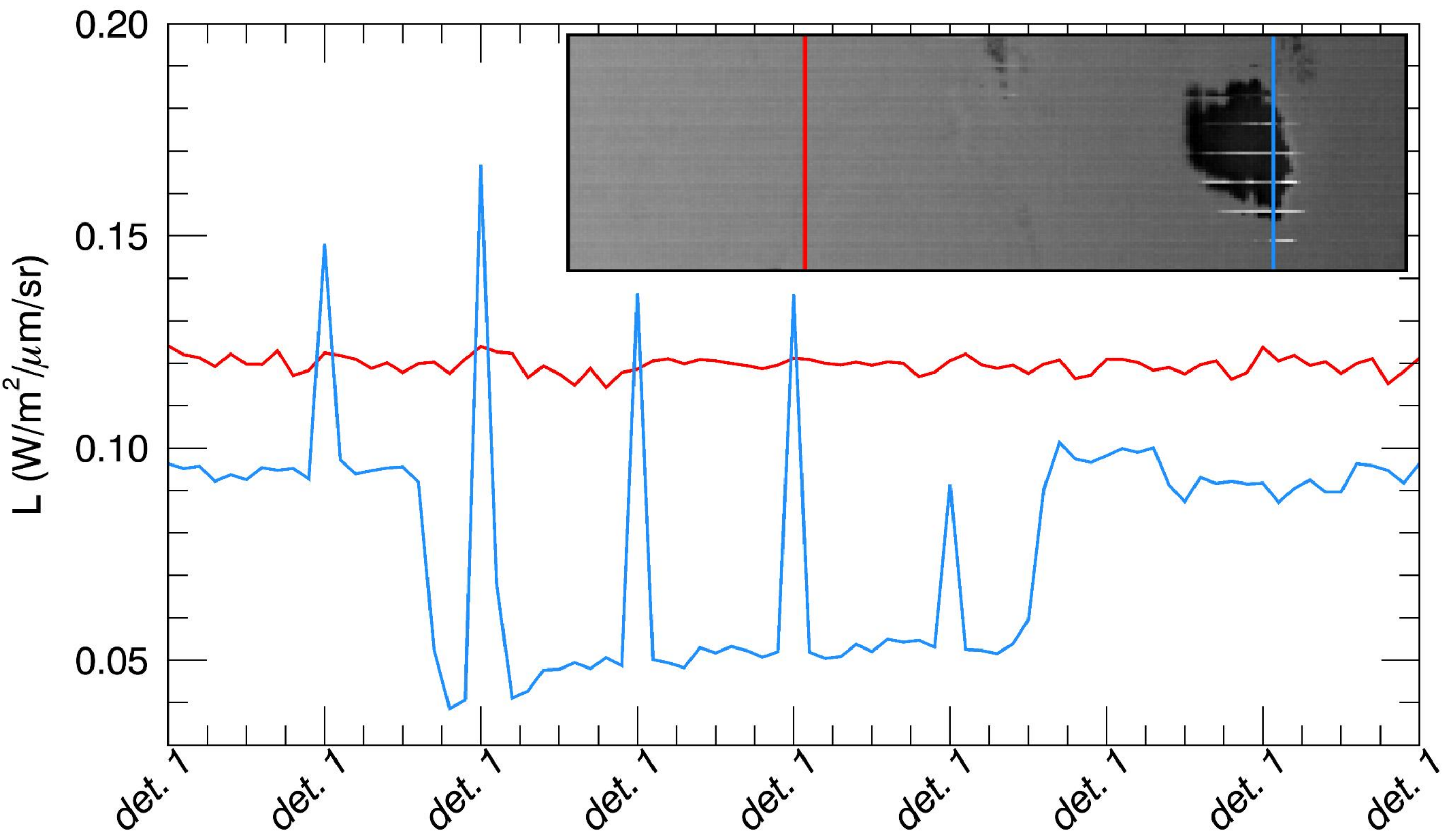}
\caption{Intensity profiles extracted from the regions marked with lines in the image on top, according to the color code. The striping artifact 
can be seen, in this sample image, affecting the darker region of the image only (however dislocated), where detector 1 outputs radiance values higher than those seen in the brighter regions of the image. }
\label{fig_profile}
\end{figure}

An analysis of the intensity profile of band 24 images across scan lines, in the track direction, illustrated in Fig. \ref{fig_profile}, 
shows detector 1 with systematically higher radiance values than the other 9 detectors, generally in dark regions of 
the images, while other regions seem much less affected.  

\begin{figure}[!t]
\centering
\includegraphics[width=3.5in,height=1.6in]{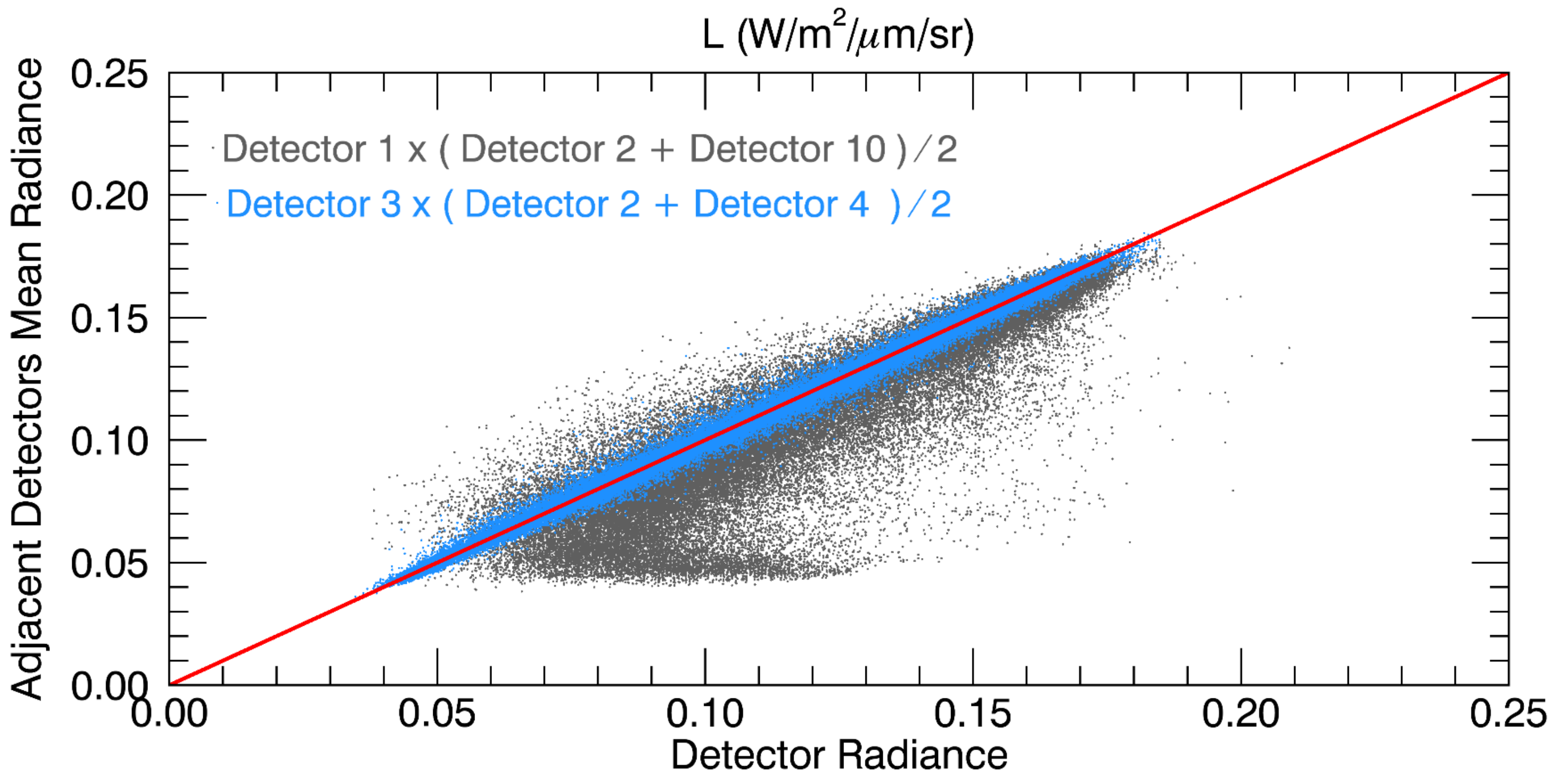}
\caption{In blue, detector 3 radiance values are compared to the average values of detectors 2 and 4, for a sample granule. The blue points follow the red $x=y$ line
closely and serve as proxy for healthy detector behavior. In contrast, the gray points correspond to detector 1 versus the average values of detectors 10 and 2, and illustrate the
out of family behavior of detector 1.}
\label{fig_neighbor}
\end{figure}

Such odd behavior can be further appreciated in Fig. \ref{fig_neighbor}, 
where we show a plot of pixel radiances taken from a sample granule for a single detector versus the average radiances from the adjacent detectors. 
In blue, we show detector 3 radiances compared to the average radiances of detectors 2 and 4. The blue points follow 
the red $x=y$ line closely and serve as proxy for healthy detector behavior. In contrast, the gray points correspond to 
detector 1 radiances versus the average radiances of detectors 10 and 2, from which we conclude that detector 1 out of 
familly behavior occurs throughout the entire radiance range, but is much more severe towards the lower radiance levels of the sample image.  

The anomaly in detector 1 is sufficiently significant to impact the band 24 lifelong brightness temperature trend over the region of Dome
Concordia (Dome C), in Antarctica. Dome C brightness temperature trends (between 210 and 240 K) are historically updated and maintained by MCST to assess calibration stability using 
average values of all detectors \cite{wenny2012evaluation}. We separated such trends into individual detectors and noticed detector 1 from band 24 consistently 
reaching temperatures 2 K higher than the other detectors, around summertime, every year during the entire mission, while behaving normally around the wintertime, 
which corresponds to polar night.

Further investigation revealed that the anomaly in the behavior of detector 1 is indeed present since the beginning of the mission in 2002 and that it is restricted to day granules. 
Another marked characteristic of the striping artifact is that despite its general behavior of closely corresponding to the darker 
regions of the image, it appears dislocated both in the track direction (or vertically) and in frame (horizontally), as seen in the image in Fig. \ref{fig_profile}, a telltale of electronic signal crosstalk.   

\subsection{Crosstalk Characterization from Lunar Images} 

Lunar observations, used in sensor on-orbit calibration, are routinely scheduled for Aqua MODIS. Spacecraft roll maneuvers 
align the Moon with the Space View port, but a sector rotation operation causes the Moon images to be stored in the Earth View data sector instead,
where the Moon can be seen in many consecutive scans, co-registered in all bands. Because images from consecutive scans overlap, we are then able to construct images from the entire Moon from single detectors. 

\begin{figure}[!t]
\centering
\includegraphics[width=1.8in,height=1.6in]{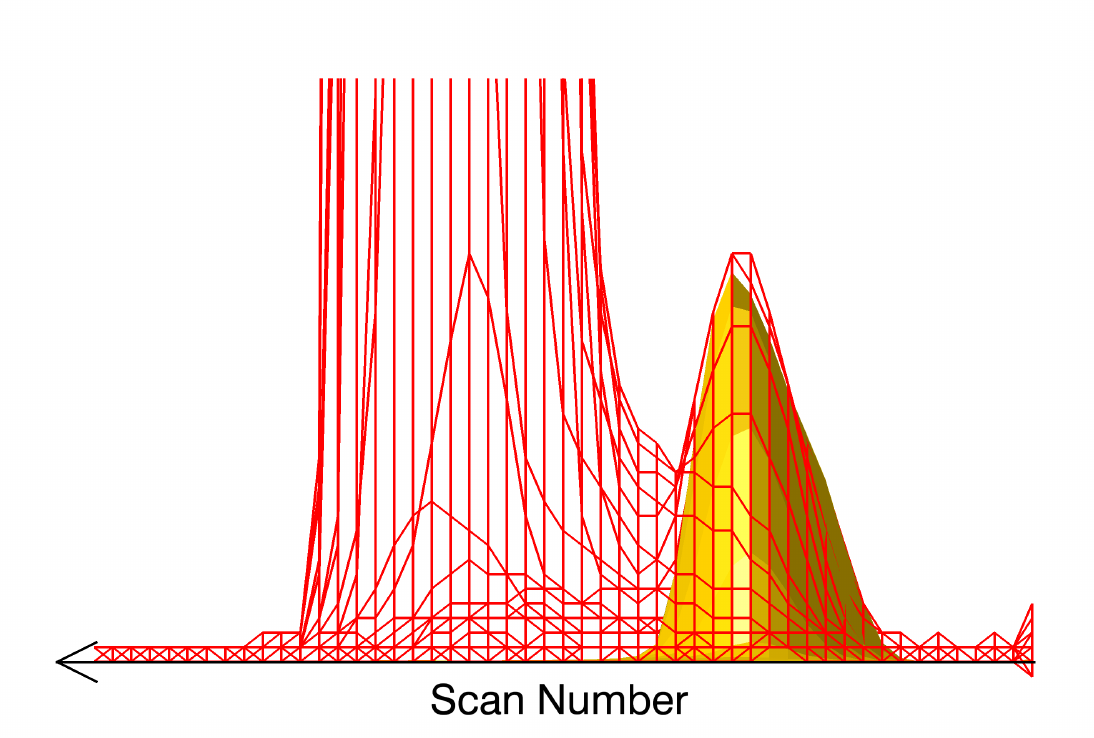}
\includegraphics[width=1.6in,height=1.5in]{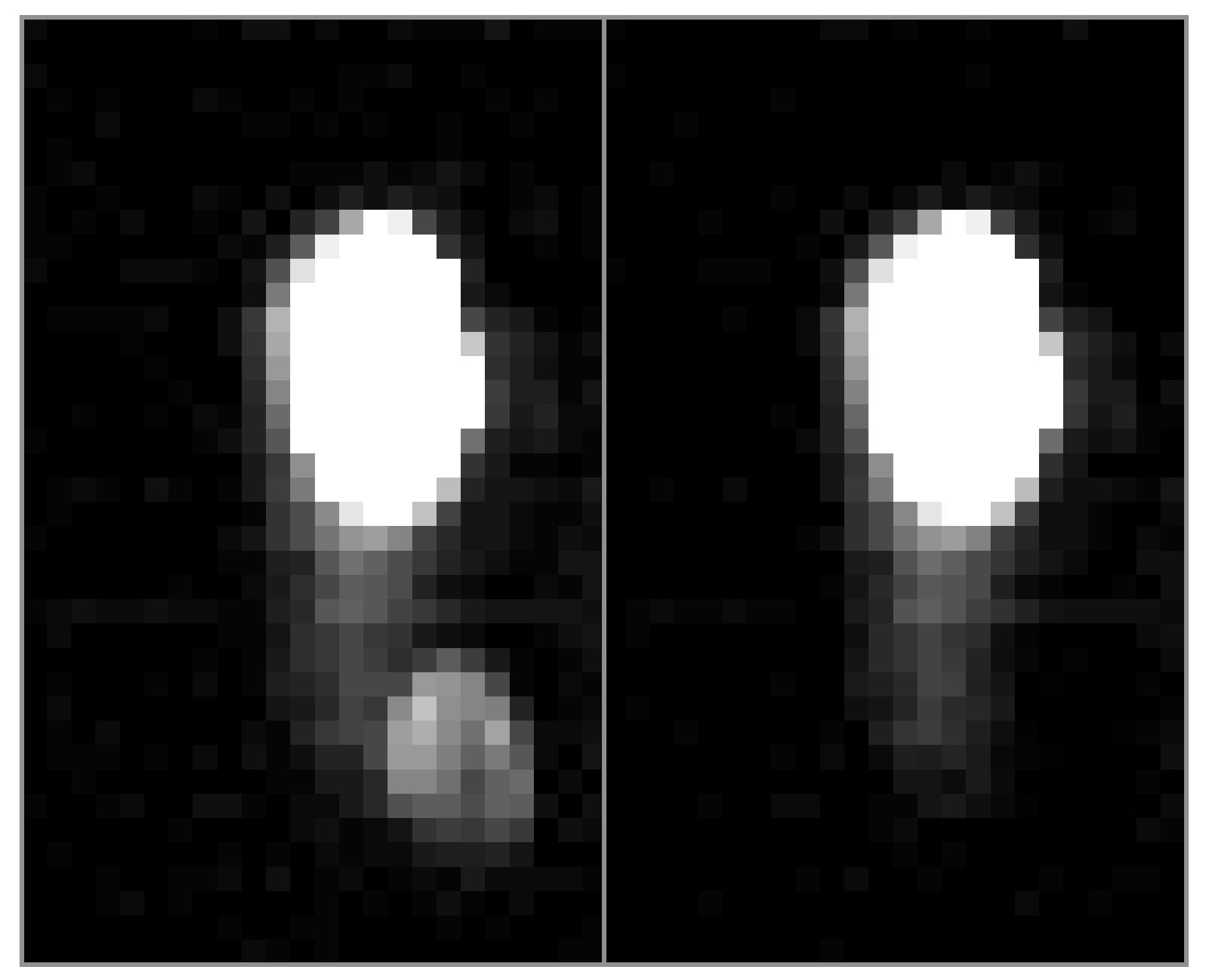}
\caption{Left panel: Moon images from a lunar event, displayed in a surface plot. While the track direction (scan number) axis points left,
the frame axis points towards the page. In red, the Moon image as seen by detector 1 from band 24 shows, on the left, the
main Moon image, truncated for clarity, and on the right, a fainter ghost. In yellow, the scaled Moon image from band 26, detector 10, aligns perfectly with the ghost from band 24.
The precise alignment and the fact that both bands are neighbors in the SWIR/MWIR FPA are strong evidence that detector 10 from band 26 is the source of the contamination. Right panel: original (left) and
crosstalk corrected (right) images of the Moon as seen by detector 1 of band 24. While the ghost image is removed from the corrected image, the tail remains intact, as desired.}
\label{fig_moon_surface}
\end{figure}

Cross-band/detector signal leaks can be readily identified in single detector images of the Moon \cite{sun2010terra,sun2015investigation,sun2015electronic,truman2016}, 
which, when illuminated by the Sun, is a bright source on both TEBs and RSBs. The left panel of Fig. \ref{fig_moon_surface} shows, in red, a surface plot 
of the Moon, viewed by band 24 detector 1. It comprehends the main image of the Moon, to the left, which is truncated for better 
visualization, and a much dimmer ghost image adjacent to it, on the right side, whose presence indicates signal contamination and which is not
present in single detector images of any of the other band 24 detectors. This absence from other detector images is strong evidence that 
this signal leak is linked to the band 24 detector 1 anomalous behavior.

The yellow surface overploted on Fig. \ref{fig_moon_surface} corresponds to the scaled Moon image as seen by band 26 detector 10, 
displaced by three frames, the exact distance between bands 24 and 26 on the SWIR/MWIR FPA. 
The precise alignment between the Moon image from band 26 detector 10 with the ghost image registered by band 24 detector 1 and the fact that 
both bands occupy the same FPA is strong evidence that detector 10 from band 26 is leaking the signal via electronic crosstalk. 
Electronic crosstalk with other bands on the same FPA, besides band 26, is possible, but would result in Moon ghosts displaced in frame number to 
different positions, according to their relative distances on the FPA to the band receiving the contaminating signal. Signal leaked by other band 26 detectors, on the 
other hand, would be displaced to a different position in scan number. In fact, 
band 24 Moon images do show other ghosts. They are, however, present in images from all the detectors in comparable intensity and are not, 
for this reason, candidates to producing the detector 1 discrepant behavior we are analyzing in this work.

\begin{figure}[!t]
\centering
\includegraphics[width=3.35in]{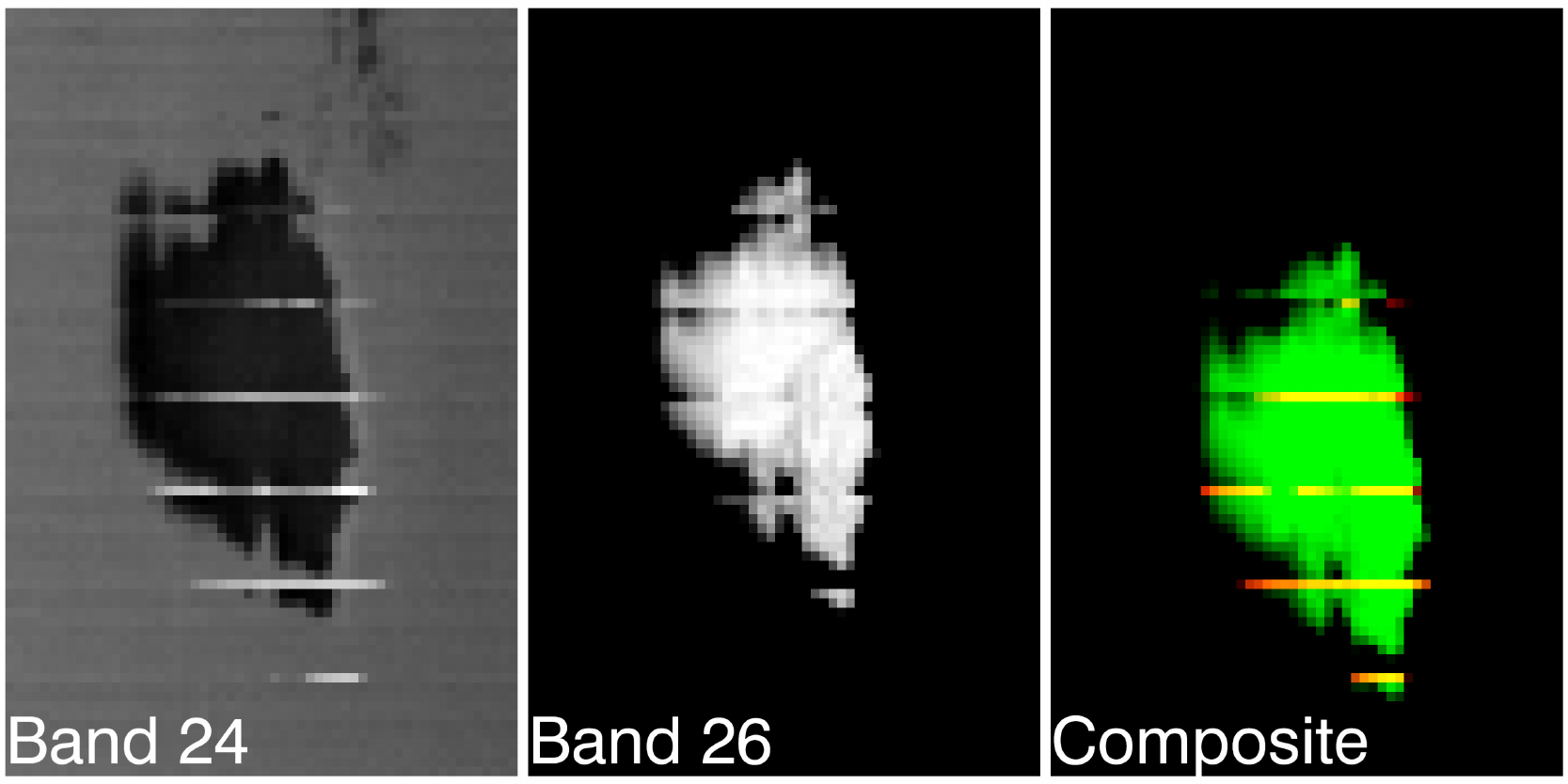}
\caption{The left and center panels show the same scene, viewed by bands 24 and 26, respectively and illustrate that bright scenes in band 26 often correspond to dark scenes in
band 24 and vice versa. The stripes in the image from band 24 appear dislocated in both the track direction and in frame,
with respect to the darker region, as is expected from electronic crosstalk contamination coming from the detector 10 of band 26. The right panel shows a RGB composite of the images from bands 24 and 26,
with the former being fed through the red channel and the latter through the green channel (no image was fed through the blue channel).
The image from Band 26 was dislocated in frame and in the track direction by the appropriate number of pixels,
which correspond to the distances between bands in the FPA and between detectors. The composite image illustrates the alignment between the stripes and the band 26 emission, once the predicted displacement is applied.}
\label{composite}
\end{figure}

The conclusion that band 26 leakage is causing the striping in band 24 imagery is further supported by the fact that band 26 is a RSB and, 
as such, registers only low intensity noise during night time, which explains our observation that only day time images of band 24 
were affected by the striping artifact. It is also interesting to notice that darker scenes in band 24 often correspond to very 
bright ones in band 26 and vice versa, as exemplified in Fig. \ref{composite}. This happens because band 26 is insensitive to land scenes and is especially sensitive to ice and clouds, regions where 
its reflective radiances are highest. Ice and clouds are cold, and thus make up dark scenes in TEBs. This explains the fact that dark regions in images from band 24 
are more severely affected by the striping in absolute values, far beyond what would be expected from the enhanced perceived contrast from contaminating signal 
on a dark background alone.

\subsection{Mitigation Strategy}

We chose to describe the signal contamination from band 26 detector 10 into band 24 detector 1 as the linear approximation 
in Equation \ref{eq_xt}, where $dn$ refers to the background subtracted instrument response, the symbol * indicates the 
contaminated signal, $c$ is the crosstalk coefficient and the subscripts indicate the sending ($s$) and the receiving ($r$) band/detector. Here, 
$S$ refers to scan number, $F$, to frame number, and $\Delta F$ to the displacement in the frame direction corresponding to the distance 
between the sending and receiving bands in the FPA.
 
\begin{equation}
dn_{r}(S,F) = dn_{r}^{*}(S,F) - c \times dn_{s}(S,F+\Delta F)  
\label{eq_xt}
\end{equation}

This same strategy has been adopted by different authors working in electronic crosstalk characterization and mitigation on Terra MODIS images 
from bands 2 and 27--30 \cite{sun2010terra,sun2015investigation,sun2015electronic,truman2016}, and recently, on the derivation of crosstalk coefficients for Aqua bands 27-30 \cite{rs8100806}, 
but had not so far been applied to Aqua MODIS image correction. 
While a long known issue in Terra, Aqua MODIS improved electronics had always been thought to have rid its Earth imagery from significant electronic
crosstalk impacts.

In order to derive the crosstalk coefficient, it is necessary to know what the uncontaminated signal from band 24 detector 1 (the left side term of equation \ref{eq_xt}) 
ought to be in some case. Single detector lunar images are the best option, in this case, because they offer an image where the contaminating 
signal is restricted to the region of the ghost, which ideally sits on a dark background. In reality that is not so due to other artifacts 
present in the image, the impact of which we worked to minimize, as described ahead.

In the particular case addressed in this work, the ghost in the image from band 24 detector 1, being produced by a detector 10 leakage, 
is as far from the main image in scan number as possible, but only slightly removed from it, by three pixels, in the frame direction due to 
the proximity between bands 24 and 26 in the SWIR/MWIR FPA. Such displacement from the main lunar image is not, however, enough to free the ghost image 
from being partialy contaminated by the intensity profile of the real Moon image. In many TEB lunar images, including that of 
band 24 detector 1, the Moon shows a tail that extends away from the main image in the track direction. 
This tail does partially overlap with our ghost image, as seen in the right panel of Fig. \ref{fig_moon_surface} and requires us to carefully select the 
regions of the ghost to be used in the derivation of the crosstalk coefficient. 

\begin{figure}[!t]
\centering
\includegraphics[width=3.5in]{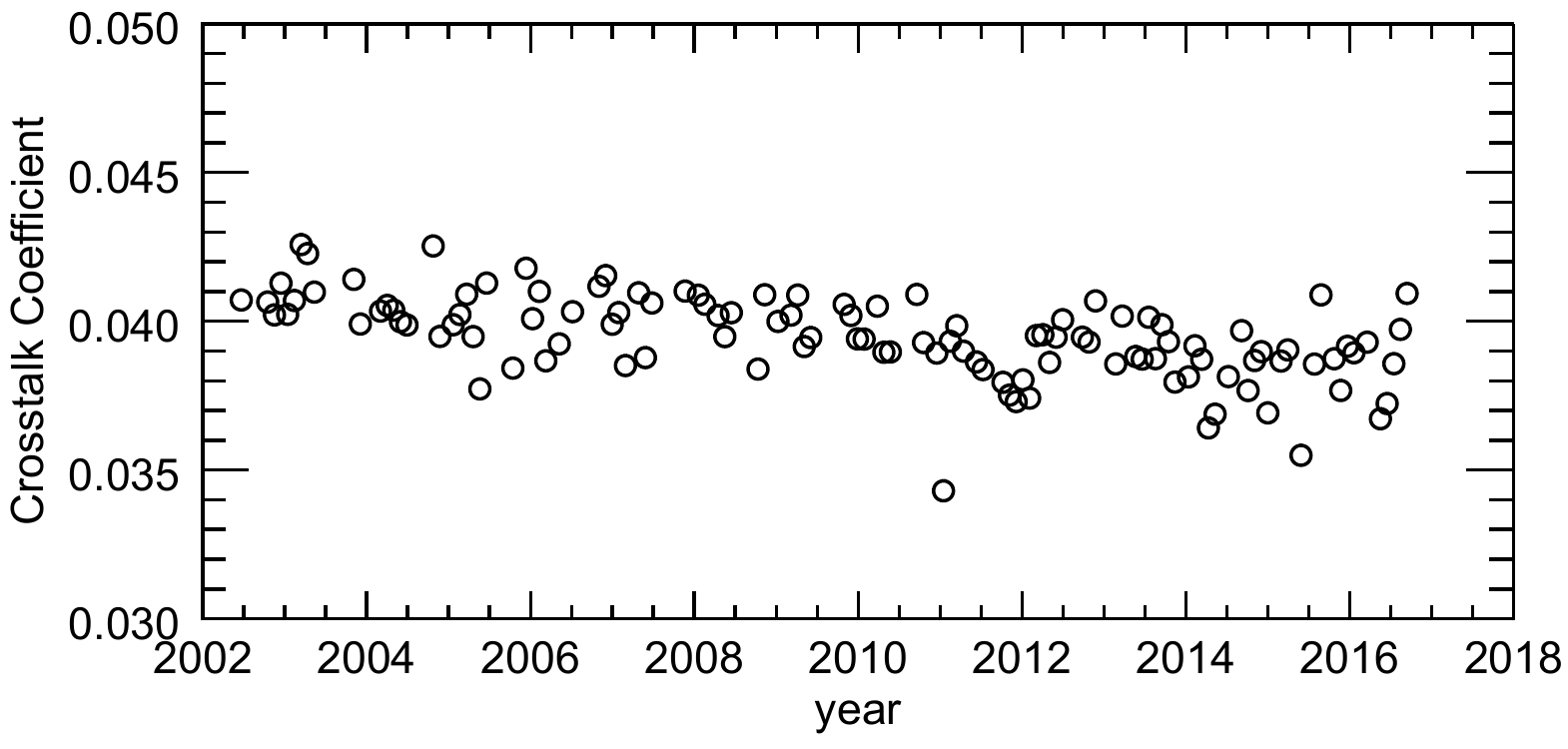}
\caption{Crosstalk coefficient trending plot. The coefficients were derived from 116 Moon images obtained from scheduled lunar events over the entire mission. It can be seen that
the general trend tends to lower $c$ values towards more recent years.}
\label{fig_trend}
\end{figure}

Portions of the ghost image that do not overlap with the tail of the main Moon image sit in a region where the counts should, in a non-crosstalk contaminated image, 
nominally be zero, apart from noise and after background subtraction. We can thus assume that all signal present in this 
region comes exclusively from the detector 10 of band 26 and derive $c$, for every pixel, from the ratio between 
$dn_{r}^{*}(S,F)$ and $dn_{s}(S,F+\Delta F)$. The final coefficient used on the correction of Earth images will then be the mean 
$c$ from all pixels considered within 1 sigma. We derived crosstalk coefficients for 116 Moon images from scheduled lunar events throughout the mission lifetime. 
Fig. \ref{fig_trend} shows $c$ has, as a general trend, been slowly decreasing with time. The derived coefficients were applied in the correction of the Moon images used to derive them, 
with the purpose of testing the ability of the adopted strategy in mitigating the artifact. An example can be seen in the right panel of Fig. \ref{fig_moon_surface}.

In this work, we did not derive new calibration coefficients used to convert the instrument
response to radiance. For MODIS TEBs, these can be derived on-orbit using a Black Body (BB).
On Aqua, it is kept at 285 K during normal operation and used to derive the linear coefficient (usually referred to as $b_{1}$)
at every scan. Every 3 months, the BB goes through a warm-up/cool-down cycle used to
derive on-orbit off-set ($a_{0}$) and non-linear ($a_{2}$) coefficients. In the specific case where the contaminating signal originates in a RSB,
we do not expect the BB images used in the derivation of the calibration coefficients to suffer a relevant impact. While IR thermal bands
view the BB as a bright source, RSBs, which are sensitive to shorter wavelengths, see it dark and register very low counts
in comparison and are not expected to be able to contribute significantly to the receiving band BB image.

\section{Results}

\begin{figure}[!t]
\centering
\includegraphics[width=3.5in,height=1.81in]{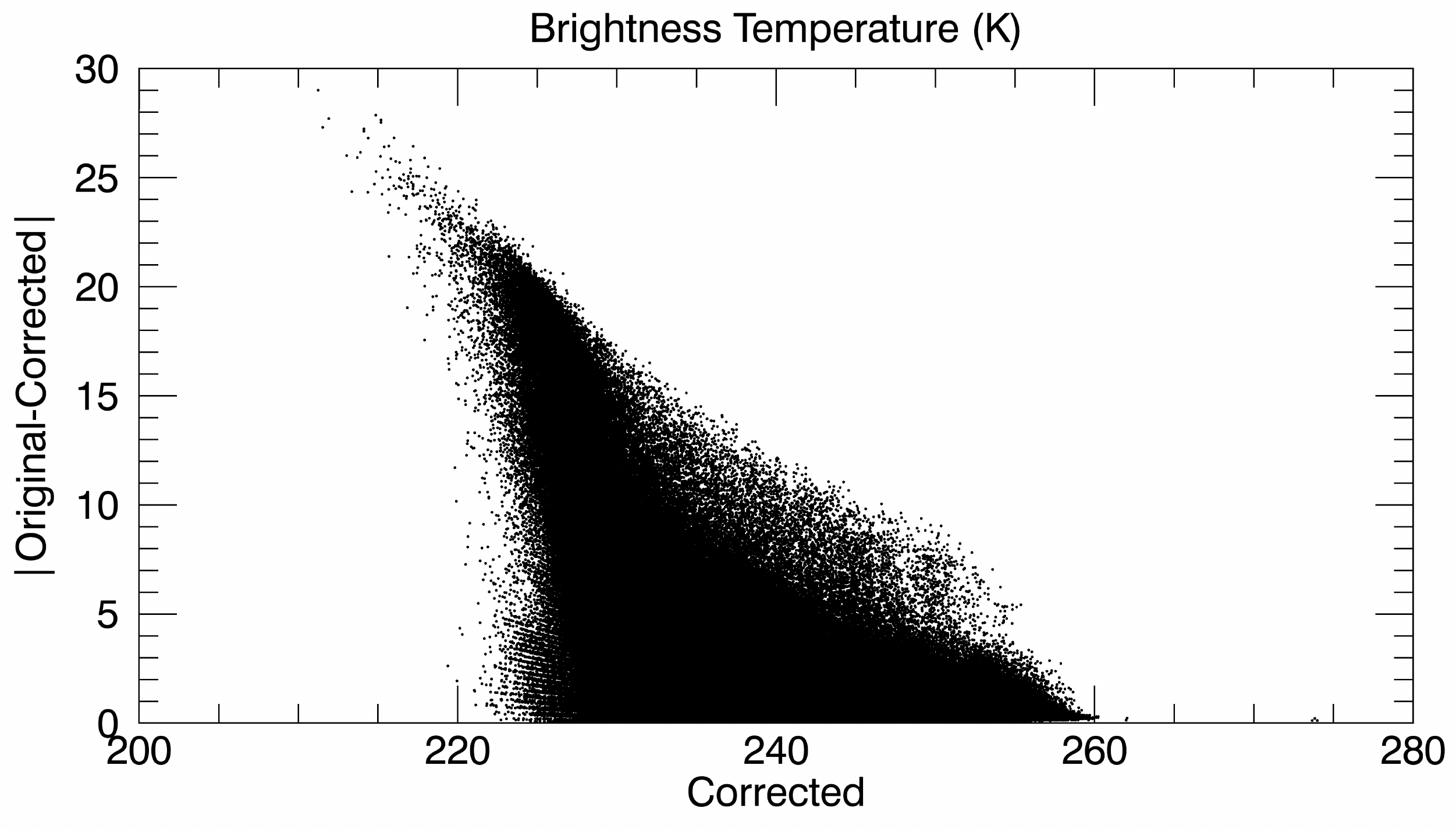}
\caption{Brightness temperature correction applied to every detector 1 pixel in band 24 images spaning 3 orbits. Only day granules were used. At T$_{typ}=250$ K, the maximum correction applyed was of 9 K.}
\label{fig_tempcorrection}
\end{figure}

\begin{figure}[!t]
\centering
\includegraphics[width=3.5in,height=1.81in]{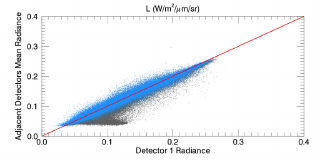}
\caption{Pixel radiances for detector 1 versus the mean radiances of neighboring detectors (detectors 10 and 2) before and after 
correction are shown as gray and blue points, respectively. Three orbits worth of day granules are shown. 
This figure illustrates the ability of the correction in restoring the behavior of detector 1, which follows the $x=y$ red line much closer after correction.}
\label{fig_neighbors_correction}
\end{figure}

\begin{figure}[!h]
\centering
\includegraphics[width=3.5in,height=2in]{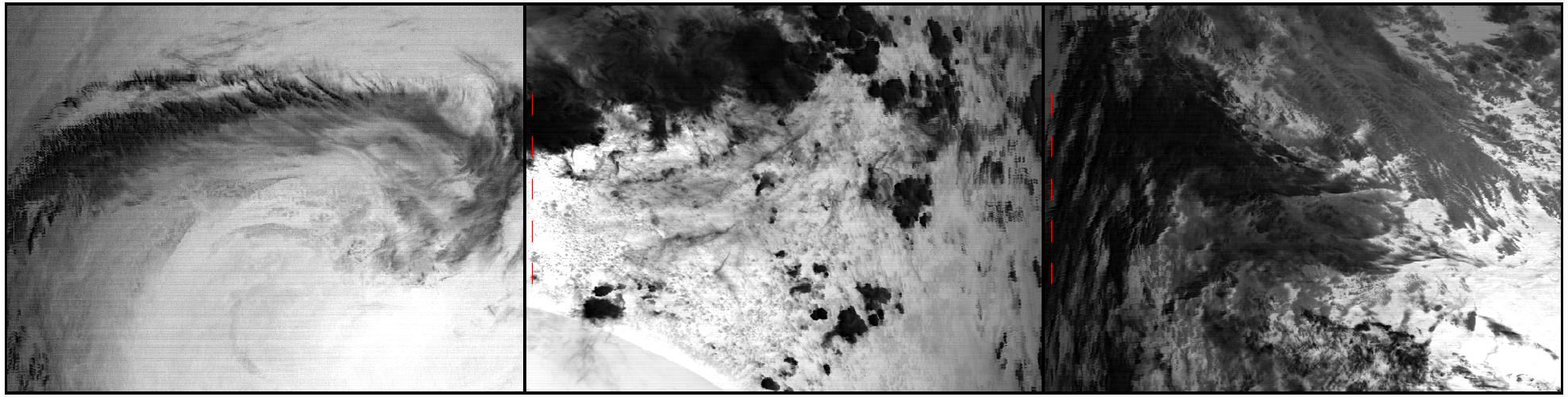}
\caption{Sample images from Fig. \ref{fig_striping}, now corrected from crosstalk contamination, illustrating the successful mitigation of the artifacts caused by detector 1 
discrepant behavior. The vertical red lines correspond to the regions used in the intensity profile plots shown on Fig. \ref{fig_profiles}. 
The grayscale colormap of each of the three images is scaled independently, but is the same as that of the corresponding uncorrected image in Fig. \ref{fig_striping}.}
\label{fig_corrected}
\end{figure}

\begin{figure}[!h]
\centering
\includegraphics[width=3.5in]{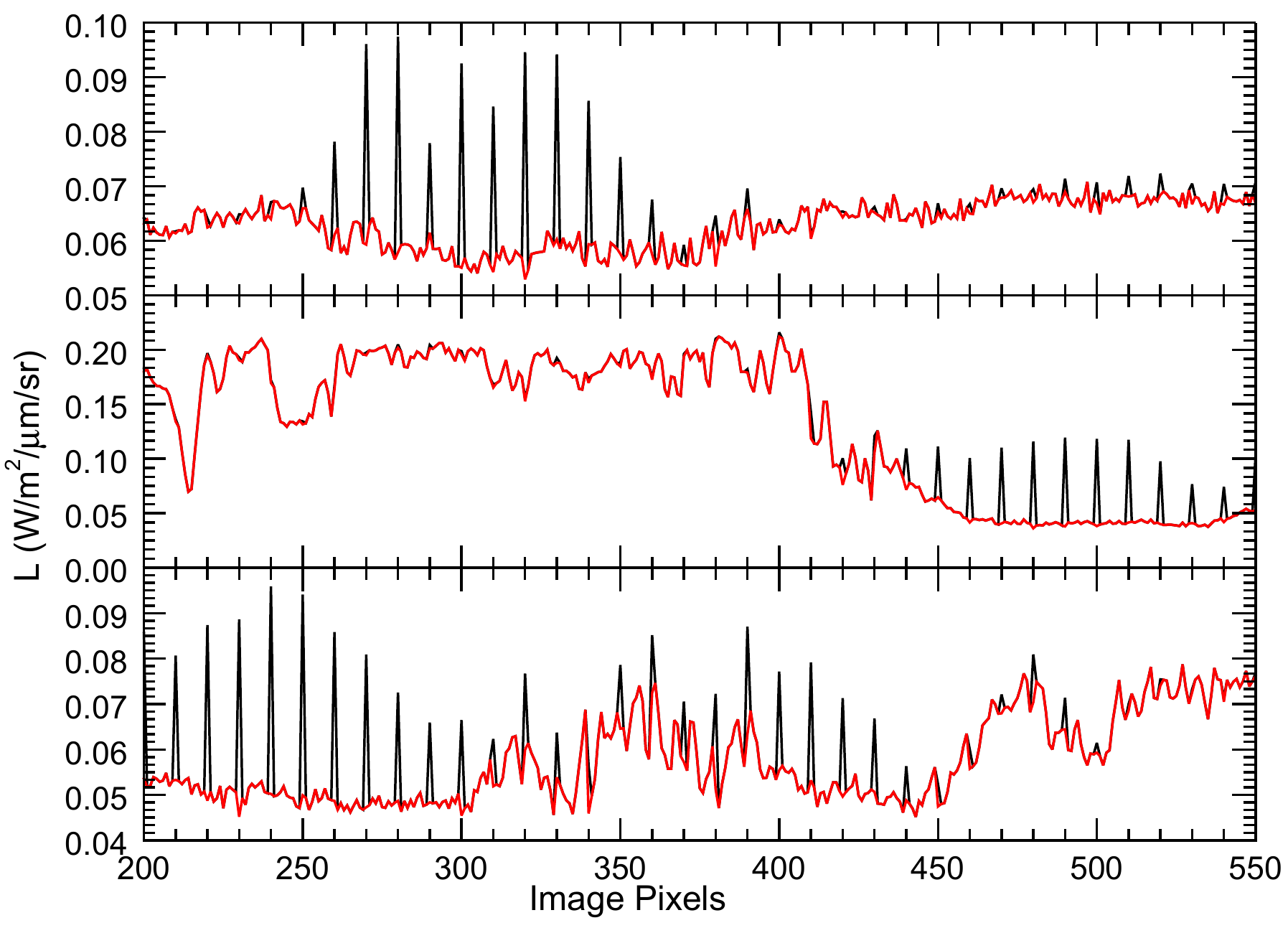}
\caption{Intensity profiles corresponding, from top to bottom, to the images in Figs. \ref{fig_striping} (black) and \ref{fig_corrected} (red), from left to right. 
The profiles were extracted from the regions marked with red vertical lines.}
\label{fig_profiles}
\end{figure}

Once the crosstalk coefficients were derived, we applied the correction to 3 orbits of consecutive L1A daytime granules as to span varied scenery, using $c$ from a lunar 
event close in time to the observations. We then produced new radiometrically 
calibrated L1B images. The maximum correction applied at T$_{typ}$ was of 9 K, a number that increases for lower temperatures, as seen in Fig. \ref{fig_tempcorrection}. 
We noticed that all but one of the granules collected showed obvious striping upon visual inspection, a fact that 
attests the prevalence of the artifact among day granules. Fig. \ref{fig_neighbors_correction} shows a plot of pixel radiances for detector 1 versus the mean radiances 
of neighboring detectors before and after correction, for all the day granules collected.  
The sample images in Fig. \ref{fig_striping} were corrected and the results are now shown in Fig. \ref{fig_corrected}, free from the striping artifact of detector 1. 
The corresponding corrected and original along track intensity profiles extracted from one column of each of the sample images are shown in Fig. \ref{fig_profiles}. 
These figures illustrate the ability of the correction applied in mitigating the striping artifact, 
restoring the anomalous behavior of detector 1 to one more ``in familly'' with the other detectors of band 24. 

The striping artifact described and mitigated in this work is not the only one affecting Aqua MODIS band 24 imagery, although it is the most severe. 
Visual inspection of the images shows further detector to detector inhomogeneities, perceived as striping, especially darker stripes on 
intermediary radiance scenes. These are, however, considerably less significant than the 
artifact addressed in this work and cannot be easily identified in the intensity profile plots shown.       

Preliminary analysis of Terra MODIS band 24 Earth and Moon 
images showed similar signs of crosstalk contamination as those described in this work, for Aqua MODIS. The impact analysis and mitigation effort concerning Terra MODIS 
band 24, as well as potential contamination in other Aqua MODIS TEBs will be addressed in a future work.   

\section{Conclusion}

In this work, we have reported on the detection of striping in Aqua MODIS band 24 images, corresponding to increased counts being output by detector 1. 
We showed the anomaly to be ubiquitous among day granules and absent from night images, relevant -- reaching a maximum impact of 9 K for T$_{typ}$ and higher for lower temperatures --, and present 
since mission beginning. We linked the artifact to the crosstalk ghost image apparent in single detector Moon images from band 24, detector 1 and absent from lunar images by all other band 24 detectors 
and identified detector 10 from band 26 as the source of the contaminating signal. Band 26 is located next to band 24 in SWIR/MWIR FPA and is a RSB, which explains the absence of striping in images from night granules.

Having chosen to describe the crosstalk contaminating signal as being proportional to the signal from the sending detector, we then used images from scheduled Moon observations throughout the mission to derive 
linear crosstalk coefficients, whose trend tends slightly downward with time. 

Finally, we applied the crosstalk correction to 3 orbits of consecutive day granules and were able to successfully mitigate the artifact originating from the discrepant behavior of detector 1 and improve the imagery quality 
and product accuracy.

\section*{Acknowledgment}

The authors would like to thank other members of MCST and in particular Dr. Truman Wilson for the fruitful discussions and suggestions to this manuscript and Amit Angal for providing us with information on MODIS' data. 
GRK made use of and acknowledges IDL libraries \textit{IDL Astronomy Users Library}\cite{1993ASPC...52..246L}, \textit{pp\_lib}\cite{paulo_penteado_2016_56247}, and \textit{mglib}.

\bibliographystyle{IEEEtran}
\bibliography{IEEEabrv,gk_refs}
\end{document}